\documentclass[nato]{crckbked}

\usepackage{graphicx}
\usepackage{klucite}

\makeindex

\begin{document}
\numreferences

\begin{article}

\begin{opening}
\title{`Taxonomy' of Electron Glasses}

% Please give the authors names and email, and put all the authors 
% from the same institution together as  in this example
\author{N.P. \surname{Armitage}\email{npa@physics.ucla.edu}}
\author{E. \surname{Helgren}}
\author{G. \surname{Gr\"{u}ner}}
\institute{Department of Physics and Astronomy, \\
University of California, Los Angeles, CA 90095}

% Authors, as they appear in the table of contents
\runningauthor{ N.P.Armitage, E.Helgren G. Gr\"{u}ner}

\begin{abstract}
We report measurements of the real and imaginary parts of the AC conductivity in the quantum limit, $ \hbar\omega > k_{B}T$ of insulating nominally uncompensated n-type silicon.  The observed frequency dependence shows evidence for a crossover from interacting Coulomb glass-like behavior at lower energies to non-interacting Fermi glass-like behavior at higher energies across a broad doping range.  The crossover is sharper than predicted and cannot be described by any existing theories.  Despite this, the measured crossover energy can be compared to the theoretically predicted Coulomb interaction energy and reasonable estimates of the localization length obtained from it.  Based on a comparison with the amorphous semiconductor NbSi, we obtain a general classification scheme for electrodynamics of electron glasses.

\end{abstract}
\end{opening}

\section{Introduction}

Strong electronic interactions are known to play a central role in disordered solids, of which Coulomb glasses are a canonical example.  The lack of metallic screening on the insulating side of the metal-insulator transition (MIT) enables long-range Coulomb interactions \cite{Helgren01}.  Efros and Shklovskii (ES), following the original considerations for the non-interacting Fermi glass case of Mott\cite{Mott}, derived a form for the T = 0 K photon assisted frequency dependent conductivity describing the crossover from interacting Coulomb glass-like behavior to Fermi glass-like behavior\cite{ES85}.  These derivations were based on a theory of resonant absorption\cite{TanakaFan} and take into account the mean Coulomb interaction between two sites forming a resonant pair $U(r_{\omega}) = e^{2}/\varepsilon_{1} r_{\omega}$, where $r_{\omega} = \xi[ln(2I_{0}/\hbar\omega)]$ is the most probable hop distance between pairs and $\varepsilon_{1}$ is the dielectric constant.  The real part of the ES crossover form for the frequency dependent conductivity is:

\begin{equation}
    \sigma_{1} = \beta 
    e^{2}g_{0}^{2}\xi^{5}\omega[ln(2I_{0}/\hbar\omega)]^{4}[\hbar\omega + 
    U(r_{w})]
    \label{eq:ESxover}
\end{equation}

where $\beta$ is a constant of order one, $g_{0}$ is the non-interacting single particle density of states (DOS), $I_{0}$ is the pre-factor of the overlap integral and $\xi$ is the localization length.  The concentration dependent localization length is predicted to diverge as $ \left(1 - \frac{x}{x_{c}}\right)^{-\nu}$ as the MIT is approached, where $x$ is the dopant concentration, $x_{c}$ is the critical dopant concentration of the MIT ($x_{c} = 3.52 \times 10^{18} cm^{-3}$ in Si:P\cite{Stupp}) and $\nu$ is the localization length exponent.

Neglecting logarithmic factors, Eq.  (\ref{eq:ESxover}) predicts a gradual crossover from linear to quadratic behavior as the incident photon energy exceeds the interaction energy of a typical charge excitation.  For the case where the photon energy, $\hbar\omega > U(r_{\omega})$, one recovers the quadratic frequency dependence, plus logarithmic corrections, that Mott originally derived for the non-interacting Fermi glass case\cite{Mott}.  In the opposite limit, $\hbar\omega < U(r_{\omega})$ the conductivity shows an approximately linear dependence on frequency, plus logarithmic corrections, and the material is called a Coulomb glass.  We should note that Eq.  (\ref{eq:ESxover}) was derived for the case where $\hbar\omega > \Delta$, the Coulomb gap width.  However a quasi-linear dependence (albeit with a different pre-factor) and an eventual crossover to Mott's non-interacting quadratic law is still expected even for the case where $\hbar\omega < \Delta$.

\begin{equation}
    \sigma_{1} \approx 
    \frac{1}{10}\frac{\omega\varepsilon_{1}}{ln(2I_{0}/\hbar\omega)}
    \label{eq:ESinGap}
\end{equation}

There is a lack of experimental evidence to either corroborate or disprove Eq.  (\ref{eq:ESxover}) due to the difficulties associated with performing frequency dependent measurements in the so-called quantum limit, i.e.  $ \hbar\omega > k_{B}T$, but at small enough photon energies so as to not be exciting charge carriers to the conduction band.  Moreover, in order to study the possible crossover from Mott to ES type behavior, one must measure across a broad enough bandwidth centered about the characteristic crossover energy scale for instance the Coulomb interaction energy $U$ or the Coulomb gap width\cite{ES75}, $\Delta$.

There have been some very recent experiments that have attempted to address these issues.  M.  Lee et al.  found that for concentrations close to the MIT the expected linear to quadratic crossover occurs, but is much sharper than predicted \cite{MLee01}.  They proposed that this sharp crossover was controlled not by the average interaction strength $U$ as in Eq.(\ref{eq:ESxover}) \cite{ES85}, but instead by a sharp feature in the density of states, i.e.  the Coulomb gap \cite{MLee99}.  They postulated that this Coulomb gap was not the single particle one measured in tunneling, but rather a smaller "dressed" or renormalized Coulomb gap that governs transport.  There is some evidence from DC transport that such a feature exists, at least close to the MIT \cite{MLee2000}.

\section{Experiment}

Nominally uncompensated n-type silicon samples were obtained from Recticon Enterprises Inc.  A Czochralski method grown boule with a phosphorous gradient along its length was cut into 1 mm thick discs.  Room temperature resistivity was measured using an ADE 6035 gauge and the dopant concentration calibrated using the Thurber scale\cite{Thurber}.  The Si:P samples discussed here span a range from 39\% to 69\%, stated as a percentage of the sample's dopant concentration to the critical concentration at the MIT.  A number of samples were measured before and after etching with a $4\% HF + 96\% HNO_{3}$ solution; this resulted in no difference in the results.

In the millimeter spectral range, 80 GHz to 1000 GHz, backward wave oscillators (BWO) were employed as coherent sources in a transmission configuration\cite {Schwartz}.  The transmitted power through the Si:P samples as a function of frequency was recorded.  For plane waves normally incident on a material, resonances occur whenever the thickness of the material is an integer number of half wavelengths.  Both components of the complex conductivity can be uniquely determined for each resonance.  The real part of the conductivity was evaluated at microwave frequencies from the measured loss of highly sensitive resonant cavities at 35 and 60 GHz via the perturbation method.  This is a common technique and is described in the literature\cite{Gruner}.  The conductivity as determined from the resonant cavity data was normalized to the DC conductivity at higher temperatures, at above approximately 25 K.  The resonant cavity data confirmed the linear dependence on frequency of the real part of the complex conductivity into the microwave regime for the samples closest to critical.

\section{Results}

\begin{figure}
\centerline{\includegraphics[width=7cm]{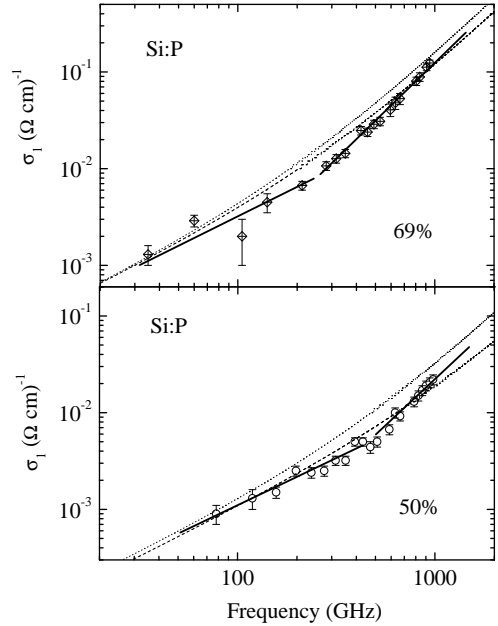}}
\caption{Low temperature frequency dependent conductivity data plotted as a function of frequency. The Si:P samples shown are at 50\% and 69\% dopant concentration relative to the critical concentration. The solid lines are linear and quadratic fits to the lower and upper portions of the data respectively. The dotted and dashed lines are fits following the form of Eq. (1) using two different methods described in the text.}
\label{ArmitageCG1}
\end{figure}

In Fig.  \ref{ArmitageCG1}, we show the T$\rightarrow$0 frequency dependent conductivity for two samples.  This data, representative of all samples in our range, shows an approximately linear dependence at low frequencies and then a sharp crossover to an approximately quadratic behavior at higher frequencies.  This is the qualitatively expected behavior from Eq.  (\ref{eq:ESxover}).  However, as seen by the overlayed fits, Eq.  (\ref{eq:ESxover}) provides only a rough guide.  The solid lines are linear and quadratic fits to the low frequency and high frequency data respectively.  The dotted line is a fit to the form of the ES crossover function achieved by summing the separately determined linear and quadratic fits.  As can be seen, the crossover between linear and quadratic portions is much more abrupt than the ES function predicts.  The dashed line is a fit using the same method as Ref.  \cite{MLee01}, namely forcing the linear portion to pass through the low frequency data, as well as the origin and leaving the pre-factor of the quadratic term as a free variable.  The fit is not satisfactory in either case.  A sharp crossover as such is observed over our entire doping range and has been observed previously in an analogous system, Si:B, for samples closer to the MIT \cite{MLee01}.  Note, that a linear dependence is seen in the imaginary part of the conductivity $\sigma_{2}$ over the whole measured frequency and doping range \cite{Helgren02}.  This is consistent with theoretical predictions \cite{Efros85}.

Because our data spans a large range of concentrations, the doping dependence of the crossover energy scale can be analyzed to see whether its dependence is consistent with other energy scales, e.g.  the Coulomb interaction energy $U$ or the Coulomb gap width $\Delta$ as $per$ Ref.  \cite{MLee01}.  Recall that the Coulomb interaction energy between two sites forming a resonant pair is $U(r_{\omega}) = \frac{e^{2}}{\varepsilon_{1} r_{\omega}}$ which is dependent on concentration via the dielectric constant (measured, but not shown) and the localization length dependent most probable hop distance.  By equating the crossover energy scale to the expected functional form for this Coulomb interaction energy we are able to determine the magnitude of the localization length and its scaling exponent.  With an appropriate pre-factor in the overlap integral \cite{Shklovskii}, $I_{0} = 10^{13} s^{-1}$ for the expression for the most probable hop distance term, $r_{\omega}$, we get a localization length dependence of $\xi \propto (1 - x/x_{c})^{-0.83}$ with a magnitude of 21.2, 19.9, 20.1, 14.5, 14.3 and 13.0 nm for the 69\%, 62\%, 56\%, 50\%, 45\% and 39\% samples respectively.  The localization length exponent is close to unity, the value originally predicted by McMillan in his scaling theory of the MIT \cite{McMillan}, and the magnitude of the localization length is reasonable.  Due to the fact that we obtain reasonable estimates for the relevant physical parameters over the whole doping range, we do not favor the previous speculation that it is in fact the Coulomb gap energy that creates the sharp crossover and hence sets its energy scale\cite{MLee01}.

The approximately linear power law seen in the Coulomb glass regime at low $\omega$ in Fig.  \ref{ArmitageCG1} of the conductivity can be expressed with the imaginary part as a simple Kramer-Kronig compatible form, $ \sigma(\omega) = A (i\omega)^{\alpha} = \ A \omega ^{\alpha} cos \left( \frac{\pi\alpha}{2} \right) +i A \omega^{\alpha} sin \left( \frac{\pi\alpha}{2} \right).$ In order to determine the power $\alpha$ one can take the ratio of $|\sigma_{2}|$ versus $\sigma_{1}$ (with the frequency as a variable).  The power $\alpha$ is given by,

\begin{equation}
    \ \alpha = \frac{2}{\pi} tan^{-1} \left( \frac{|\sigma_{2}|}{\sigma_{1}} 
    \right).
    \label{eq:alpha}
\end{equation}

\begin{figure}
\centerline{\includegraphics[width=7cm]{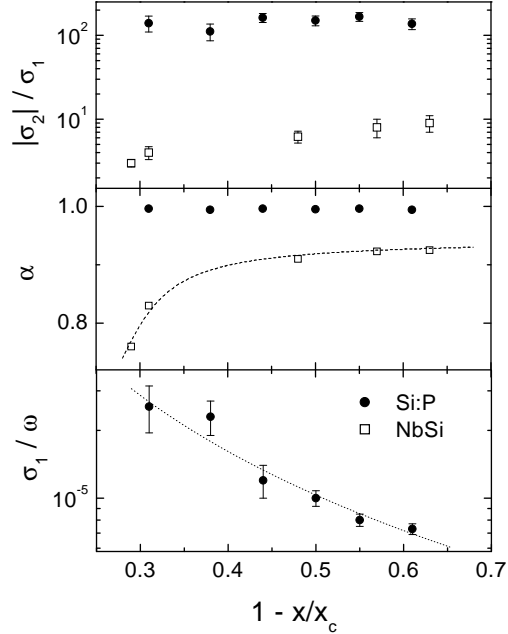}}
\caption{The upper panel shows the ratios of the imaginary to the real part of the complex conductivity for samples of Si:P and amorphous NbSi.  The NbSi data is adapted from Ref 1.  The middle panel shows the calculated powers of $\alpha$ as determined from Eq.  3.  The dashed line through the NbSi data is a guide to the eye.  The bottom panel shows the divergence of the prefactor of the real part of the conductivity, and the dotted line is a simple power law fit.}
\label{ArmitageCG2}
\end{figure}

Fig.  \ref{ArmitageCG2}a shows the ratio mentioned above of the imaginary to the real part of the dielectric constant for Si:P.  Similar data from amorphous NbSi is included for comparison purposes\cite{Helgren01}.  We note that this ratio for Si:P is large and essentially constant across the entire doping range.  From Eq.  (\ref{eq:ESinGap}), one expects $|\sigma_{1}|$ to be approximately equal to $|\sigma_{2}|$ to within a factor of 2-5 (with a reasonable estimate for $I_{0}$) because $|\sigma_{2}| \propto \varepsilon_{1} \cdot \omega$.  Applied to Si:P, the theory correctly predict a linear correspondence between $\sigma_{1}$ and $|\sigma_{2}|$, but incorrectly predicts the proportionality by at least a factor of thirty.  The proportionality is near the predicted value for NbSi, but has a dependence on the doping concentration which is presumably due to entering the quantum critical (QC) regime as discussed below.  Here we have used the susceptibility $4 \pi \chi$ of the dopant electrons (i.e. with the background dielectric constant of silicon subtracted) in the expression for the magnitude of the imaginary component of the conductivity in Eq. \ref{eq:alpha}.

The middle panel in Fig.  \ref{ArmitageCG2} shows the power $\alpha$ as determined by Eq.  (\ref{eq:alpha}).  The values for Si:P are approximately equal to, but slightly less than one, consistent with Fig.  \ref{ArmitageCG1}.  This indicates that the prefactor of the real and imaginary components of the complex conductivity have the same concentration dependence.  The situation is different for NbSi.  Near the MIT, $\sigma(\omega)$ is expected to cross over to the QC dynamics \cite{Carini98,Henderson}, i.e.  $\sigma_{1} \propto \omega^{1/2}$ when $\xi$, the localization length, is of the same scale as $\ell_{\omega}$, the dephasing length (the characteristic frequency dependent length scale) \cite{Sondhi}.  This should be a smooth crossover and therefore looking at a fixed window of frequencies, a continuous change from $\omega \rightarrow \omega^{1/2}$ is expected, similar to that measured for NbSi shown in the middle panel of Fig \ref{ArmitageCG2}.

Setting the relations for localization length and dephasing length equal \cite{Sondhi}, one finds the crossover condition for the frequency in terms of the normalized concentration, $\omega \propto \ell_{0}^{z}\left(1-\frac{x}{x_{c}}\right)^{z\nu}$ where $z$ is the dynamic exponent.  As the prefactor can vary from system to system, the fact that we see an $\alpha \approx 1$ across our entire doping range in Si:P, but an $\alpha$ that approaches $1/2$ in NbSi indicates that the critical regime in Si:P is much narrower and out of our experimental window.  This is consistent with simple dimensional arguments\cite{Kivelson} that show the crossover should be inversely proportional to the dopant density of states.  The much smaller dopant density in Si:P vs. NbSi (a factor of $10^{3}$) is consistent with a narrower QC regime as compared to NbSi.

The bottom panel in Fig.  \ref{ArmitageCG2} shows magnitude of the real part of the conductivity as the MIT is approached.  This demonstrates that the prefactor A can be written as a function of the normalized concentration, i.e.  $A \propto f(1-\frac{x}{x_{c}})$ for Si:P.  

\section{Discussion}

In $typical$ interacting systems, the effects of correlations become simpler as one goes to lower energies and/or lower temperatures.  The canonical example of this is a Fermi liquid where at T=0 and $\omega$=0 one recovers the non-interacting theory, but with parameters that are substantially modified (renormalized) from the free electron ones.  The Coulomb glass is a fundamentally important example in solid state physics because it belongs to a class of systems where this does not occur and the non-interacting functional forms are not recovered at asymptotically low energies.

As predicted in the crossover function Eq.  \ref{eq:ESxover}, we have observed in Si:P that the non-interacting functional form is recovered in the high-frequency limit and the low-frequency response shows interactions.  Within the theory this is a result of the additional internal excitation structure of a resonant pair caused by interactions and the fact that any one pair can be thought of as a distinct entity, well separated in energy from other spatially nearby pairs because of the large disorder induced energy spread.  This internal structure enables the excitation of pairs relatively deep within the Fermi sea and changes the factor in the initial state phase space from $\omega$ to $U(r) + \omega$.  In contrast, in amorphous NbSi a smooth crossover is observed as the MIT is approached from a linear frequency dependence to a power law characterized by an exponent smaller than unity.  This is consistent with the eventual frequency dependence $\sigma \propto \omega^{1/2}$ expected from quantum critical scaling arguments.  No crossover to $\omega^{2}$ was observed in NbSi.

\begin{figure}
\centerline{\includegraphics[width=7cm]{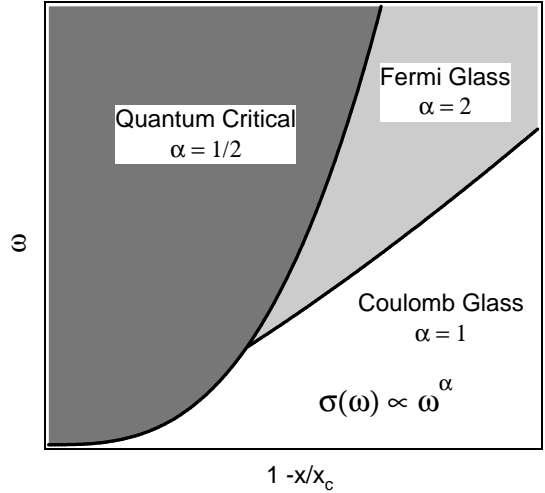}}
\caption{A schematic showing the parameter space for values taken by $\alpha$ for $\sigma \propto \omega^{\alpha}$.  Note that the boundaries drawn on the plot are smooth crossovers and not sharp onsets.  A classification based on $\alpha$ gives a taxonomy for the electrodynamics of electron glasses.}
\label{ArmitageCG3}
\end{figure}

Since it is the exponent $\alpha$ that distills the important physics (it indicates the phase space of initial states for $\alpha$ = 1 or 2 and the presence of critical dynamics for $\alpha=1/2$), we propose that one can classify the electrodynamics of electron glasses based on their $\alpha$ value.  A schematic showing the parameter space for $\alpha$ is shown in Fig.  \ref{ArmitageCG3}.  Here one has $\alpha$'s close to $1/2$ near the MIT.  There is a smooth crossover to Coulomb glass-like $\alpha=1$ at lower doping levels and an intervening non-interacting Fermi glass regime at even higher energies and lower dopings.  We expect that these general considerations are valid, despite the fact that some of the parameter space may not be accessible in certain systems.  For instance, one may begin to excite structural modes at energies high enough to see $\omega^{2}$ behavior in NbSi.  For dopings not close enough to critical in Si:P, excitation to the conduction band may be observed before critical dynamics are.

\section{Conclusion}

In summary, we have observed a crossover in the frequency dependence of the conductivity from Coulomb glass-like behavior to Fermi glass-like behavior across our entire range of doping concentrations in Si:P.  The existence of a crossover is consistent with theoretical predictions, but it is sharper than predicted.  The fact that we see the same functional form over the whole doping range (even deep into the insulating regime, where Eq.  \ref{eq:ESxover} is expected to be more valid) shows that the nature of the low energy charge excitations is qualitatively the same over the whole doping range; the inadequacy of Eq.  \ref{eq:ESxover} in describing the frequency dependent conductivity quantitatively is not limited to concentrations close to critical.  In the amorphous semiconductor NbSi, we observe a gradual crossover from $\alpha=1$ Coulomb glass-like behavior to quantum critical-like dynamics.  This comparison allows us to obtain a general classification scheme for the electrodynamics of electron glasses based on the exponent of the frequency dependence $\alpha$.  We expect that this classification or `taxonomy' will be valid even when certain regimes are not experimentally accessible.

\begin{acknowledgements}

We wish to thank Phu Tran for assisting with the cavity measurements and Barakat Alavi for assisting with the sample preparation.  We would also like to thank Steve Kivelson and Boris Shklovskii for helpful conversations.  This research was supported by the National Science Foundation grant DMR-0102405.

\end{acknowledgements}

\end{article}

\end{document}